\documentclass[referee,a4paper,12pt,traditabstract]{swsc}

\usepackage{graphicx}
\usepackage{txfonts}
\usepackage{subfigure}
\usepackage{epstopdf}
\usepackage{lineno}
\usepackage[authoryear,round]{natbib}
\usepackage[backref]{hyperref}
\usepackage{url}

 \usepackage{changes}

\hypersetup{colorlinks=true,citecolor=cyan,urlcolor=cyan,linkcolor=blue}

\begin{document}


   \title{A time dependent relation between EUV solar flare light-curves from lines with differing formation temperatures}

   
   \titlerunning{EUV Line Flare Relation}

   \authorrunning{Thiemann, Eparvier and Woods}

   \author{E. M. B. Thiemann \thanks{Corresponding Author}
          \and
          F. G. Eparvier
          \and
          T. N. Woods
          }

   \institute{Laboratory for Atmospheric and Space Physics (LASP), University of Colorado,
              3665 Discovery Drive Boulder\\
              \email{\href{mailto:thiemann@lasp.colorado.edut}{thiemann@lasp.colorado.edu}}
             }


 
  \abstract
   {Extreme ultraviolet (EUV) solar flare emissions evolve in time as the emitting plasma heats and then cools. Although accurately modeling this evolution has been historically difficult, especially for empirical relationships, it is important for understanding processes at the Sun, as well as for their influence on planetary atmospheres.  With a goal to improve empirical flare models, a new simple empirical expression is derived to predict how cool emissions evolve based on the evolution of a hotter emission.  This technique is initially developed by studying 12 flares in detail observed by the EUV Variability Experiment (EVE) onboard the Solar Dynamics Observatory (SDO).  Then, over 1100 flares observed by EVE are analyzed to validate these relationships. The Cargill and Enthalpy Based Thermal Evolution of Loops (EBTEL) flare cooling models are used to show that this empirical relationship implies the energy radiated by a population of hotter formed ions is approximately proportional to the energy exciting a population of cooler formed ions emitting when the peak formation temperatures of the two lines are up to 72\% of each other and above 2 MK.  These results have practical implications for improving flare irradiance empirical modeling and for identifying key emission lines for future monitoring of flares for space weather operations; and also provide insight into the cooling processes of flare plasma.   
   }        


   \maketitle

\section{Introduction}\label{sec:intro}
Solar flare extreme ultraviolet (EUV, 10-121 nm) emissions have highly varying time histories or light curves. Understanding, specifying, and predicting these light curves not only provides information about solar processes but also provides critical information on drivers of variability in planetary ionospheres and thermospheres.  
Solar flares begin with magnetic reconnection high in the solar corona that directly heats the magnetically confined plasma and accelerates particles along magnetic field lines away from the reconnection site.  The downward moving particles are decelerated in the dense plasma of the lower solar atmosphere, resulting in rapid heating of this relatively cool plasma.  This heated plasma subsequently flows upward along the field lines and cools by both conducting heat downward through the field line foot points and radiating energy into space.  

Aspects of flare evolution are evident in typical EUV flare light curves, where cooler-forming emission lines tend to show an early impulsive peak corresponding with the rapid heating of lower-atmosphere plasma, and hotter-forming emission lines tend to show a later more gradual peak corresponding with the plasma cooling through a line's temperature of peak formation.  These two distinct phases are often referred to in the literature as the impulsive and gradual phases, respectively.  This paper establishes the empirical relationship between gradual phase emissions from lines of differing formation temperatures.  As such, it is important to emphasize that since the gradual phase occurs as the flare plasma cools,  the emission intensity of a line increases with time as the (decreasing) temperature approaches the line's peak formation temperature, reaching a maximum at the peak formation temperature, followed by a decrease in emission intensity as the temperature decreases away from the peak formation temperature.  This causes emission lines with cooler formation temperatures to typically peak after the hotter emission lines during the gradual phase.  This mechanism also broadens the light curves of cooler emission lines with respect to hotter emission lines because the cooling rate typically decreases with time.

At Earth and other planets, solar flares cause heating and ionization in the upper atmosphere that can disrupt telecommunication and navigation systems, and increase satellite drag.  Thermospheric density and temperature enhancements, which can lead to increased satellite drag, are attributed to flare EUV emissions.  This is due to major atmospheric species having larger ionization cross-sections at longer wavelengths (but still short-ward of the ionization threshhold), and hence energy at EUV wavelengths is absorbed at higher altitudes than soft x-ray wavelengths (\cite{qian2010location}; \cite{le2012analysis}).  Thermospheric enhancements peak multiple hours after the solar flare soft x-ray and EUV maxima (\cite{qian2011variability}, \cite{sutton2006neutral}).  As such, short term forecasts of thermospheric flare response are feasible by coupling thermospheric models with accurate solar flare spectral irradiance estimates.  These spectral irradiance estimates can come from direct measurements, such as those made by the EUV Variability Experiment (EVE) onboard the Solar Dynamics Observatory (SDO) (\cite{woods2012extreme}).  However, current and future operational space weather instruments such as those on the current Geostationary Operational Environmental Satellites (GOES), as well as the recently launched GOES-R series EUV and X-ray Irradiance Sensors (EXIS) (\cite{Chamberlin2009}; \cite{Eparvier2009}), measure only a small number of select EUV bands.  Because of this unavailability of high time cadence and spectrally resolved EUV measurements, models are needed to estimate the EUV spectral irradiance as a function of time during solar flares.  

The primary goal of this paper is to present a new relationship that has the potential for improving empirical solar flare EUV irradiance models. Because solar EUV irradiance models can be broadly categorized as being either empirical or physics-based, these two model categories are briefly reviewed to provide context for this paper's primary result.  Although the topic of this paper relates specifically to flare irradiance, flare irradiance empirical modeling is still in its infancy.  As such, daily irradiance models are also discussed.  

Empirical irradiance models estimate solar irradiance by using established relationships between past spectral irradiance measurements and available irradiance proxies such as F10.7.  The  general method involves first decomposing both the spectral irradiance measurements and proxies into short and long time-scale components.  Then regression coefficients are found between the proxy and spectral irradiance values separately for each short or long time-scale component (e.g. \cite{hinteregger1981observational}, \cite{richards1994euvac}, \cite{tobiska2000solar2000}, \cite{chamberlin2007flare}, \cite{thiemann2017maven}).  Empirical irradiance models allow spectral EUV irradiance measurements made by a relatively short lived satellite or rocket mission to be extended to any period when the proxy measurements are available.  For example, F10.7 driven irradiance models can estimate spectral irradiance made back to 1947, when near continuous daily F10.7 measurements began.  

In contrast to the multitude of empirical daily averaged irradiance models, few empirical flare irradiance models are reported in the literature.   The model that has seen the most widespread use is the Flare Irradiance Spectral Model (FISM) (\cite{chamberlin2008flare}).  To model flare irradiance using commonly available flare measurements, FISM uses two well-known effects: flare limb darkening (a manifestation of solar limb darkening as discussed by e.g. \cite{pierce1977solar} and \cite{hestroffer1998wavelength}) and the Neupert Effect (\cite{neupert1968comparison}), which is the empirical linear relationship between impulsive phase flare emissions and the time-derivative of gradual phase emissions that is observed in many flares.  

Alternatively, physics-based irradiance models estimate solar irradiance by first using hydrodynamic or magnetohydrodynamic theory to model the temperature, density, abundance and magnetic field of the solar atmosphere.  Next, the spectral radiance of the individual loops are modeled from these plasma parameters using the various theories of radiation process in the solar atmosphere (e.g. atomic line and bremsstrahlung emission theory).  Observations show that quiescent coronal loops are composed of many individual strands that are heated independently, resulting in a multi-thermal loop that must be parameterized with a Differential Emission Measure (DEM) for temperature analysis.  As such, models of the quiescent corona must treat a loop as being composed of many strands with differing temperatures ( \cite{warren2002hydrodynamic}, \cite{bradshaw2003self},  \cite{reale2014coronal}).  For flare loops, observations show that they can be approximated as being isothermal and parameterized with the more simple Emission Measure (EM) in some cases (e.g. \cite{raftery2009multi}, \cite{ryan2013decay}); but in other cases, such as for the specific case of long duration flares, flare loops can be significantly multi-thermal (e.g. \cite{warren2006multithread}, \cite{warren2013observations}).   There is no consensus  on what fraction of flares are predominantly isothermal, and the isothermal approximation is still frequently used because of its simplicity.  An example of a physics-based model that has seen widespread use in the literature is the Enthalpy Based Thermal Evolution of Loops (EBTEL) model (\cite{klimchuk2008highly}, \cite{cargill2012enthalpy}), which can efficiently model average loop parameters (e.g. temperature and density) without calculating the spatial loop evolution, and this efficiency makes it an appealing option for physics-based flare irradiance modeling for space weather operations.  

For flare irradiance models used in high time cadence upper planetary atmosphere studies, empirical models, and FISM in particular, are the most widely used type of model because they depend on a small number of parameters and are computationally inexpensive.  A major deficiency of FISM for modeling EUV flare irradiance is that it assumes the time evolution of the impulsive and gradual phases is the same for all emission lines, where in reality the time evolution changes as the temperature and density of the emitting plasma evolves.  FISM models the impulsive phase from the derivative of soft x-ray measurements, applying the Neupert Effect, and assumes that the gradual phase for all EUV emissions evolves proportionally to the GOES X-Ray Sensor (XRS) 0.1-0.8 nm (-Long) band (e.g. \cite{bornmann1996goes}).  As such, FISM does not correct for the time-lag, nor the overall broadening of relatively cool EUV flare emissions with respect to the hottest EUV and SXR flare emissions.  The implication being that FISM under-predicts both the flare duration and total deposited energy.  It follows that thermospheric models driven by irradiance models that underestimate the EUV gradual phase will underestimate the corresponding temperature and density enhancements (\cite{qian2011variability}).  
 
To address this deficiency, methods that accurately estimate the time evolution of the EUV gradual phase are needed.  Physics-based irradiance models can potentially address this issue, and the EBTEL model is particularly appealing given its computational efficiency.  A number of studies have used EBTEL to predict EUV flare light curves (\cite{raftery2009multi}, \cite{hock2012thesis}, \cite{qiu2012heating}, \cite{zeng2014flare}, \cite{li2014heating}, and \cite{li2012analysis}).  However, there is typically poor agreement of modeled versus measured EUV light curves in these studies (see for example: Figures 3 and 4d in  \cite{raftery2009multi}, Figure 5 in \cite{qiu2012heating}, Figure 8 in \cite{zeng2014flare}, Figures 5 and 6 in \cite{li2014heating}, and Figure 4 in  \cite{li2012analysis}).  This paper presents an empirical approach to modeling EUV light curves a new effect that relates EUV light curves from lines with differing formation temperatures akin to how the Neupert Effect relates the gradual and impulsive phase emissions, and demonstrates that this effect can be used to accurately predict the time evolution of cooler emission lines from the light curve of a single hot emission line using a 0-dimensional (0D) equation constrained by a single parameter.  

This study begins with the observation that differences between flare light curves from hotter and cooler EUV emission lines are strikingly similar to the differences from inward and outward heat functions of a simple Lumped Element Thermal Model (LETM), frequently used in heat transport analysis.  It is then shown that the well known Cargill Model for flare cooling can be rearranged into the LETM equation, and this expression is used to motivate an empirical relationship, simply referred to as LETM, between flare light curves from hotter and cooler forming emission lines.  12 flares and 3 emission lines are examined in detail showing good agreement between the empirical LETM predictions and observations.  The analysis is then expanded to over 1100 flares and 11 emission lines, showing the empirical relation is statistically significant for emission lines with formation temperatures above approximately 2 MK.  Because LETM is highly constrained, depending only on one free parameter, and used for heat transport analysis, the possible underlying physics of the relationship are explored. 
\section{Methods}\label{sec:methods}

The literature is abound with comparisons of solar flare EUV light curves from emissions with different formation temperatures (e.g. \cite{aschwanden2001flare}, \cite{woods2011new}, \cite{chamberlin2012thermal}, \cite{ryan2013decay} ) that reflect a common trend of the cooler emission intensity peak being both delayed and broadened when compared to the hotter emission peak--features that any empirical model should capture.  This delay and broadening is qualitatively similar to how a system that both stores and dissipates thermal energy modulates heat as it flows through it (e.g. \cite{ghosh_heat_trans}).  Such systems are characterized by a LETM, where the LETM equation is

\begin{equation}
Q_{in}-\tau\frac{dQ_{out}}{dt}=Q_{out}. \label{eqn_lpfe}
\end{equation}

\noindent{}Here, $Q_{in}$ and $Q_{out}$ are the inward and outward flowing heat, respectively, and $\tau$ is the system time constant determined by the product of the system's effective thermal resistance and heat capacitance. Equation (\ref{eqn_lpfe}) can be solved numerically with the recursive equation (e.g. \cite{smith1997dsp}),

\begin{equation}
Q_{out}[n]=\left(1-e^{-1/ \tau_n}\right)\cdot Q_{in}[n] + e^{-1/ \tau_n}\cdot Q_{in}[n-1], \label{eqn_lpfe_num}
\end{equation}

\noindent{}where the temporal grid spacing must be uniform, $n$ is the time index, and $\tau_n$ is the numerical time-constant and has the same units as the grid spacing of $Q_{in}$.

In order to relate the LETM equation to cooling flare plasma, we begin with the simple 0D Cargill Model for flare cooling summarized by \cite{cargill1995cooling}.  This model assumes that conductive cooling dominates early in the cooling phase of a 
flare loop, after which radiative cooling dominates with a time constant typically on the order of tens of minutes.  Neglecting conductive cooling, a post-flare plasma loop will cool according to

\begin{equation}
Q_{in}-C\frac{dT}{dt}=Q_{out,r},
\end{equation}

\noindent{}where $Q_{in}$ is the energy flux entering the loop, $C$ is the loop constant-volume heat capacity, $T$ is the loop temperature and $Q_{out,r}$ is the radiant energy flux.  Appendix \ref{ap:a} shows that $T$ can be replaced with the product, $R_{rad}Q_{rad}$, where $R_{rad}$ is the thermal resistance for radiation.  This substitution results in,  
\begin{equation}
Q_{in}-C\frac{dR_{rad}Q_{out,r}}{dt}=Q_{out,r}. \label{eq:unfactored}
\end{equation}

$R_{rad}$ can be approximated as being quasi-static with respect to $Q_{out,r}$ when

\begin{equation}  
\left| R_{rad} \frac{dQ_{out,r}}{dt} \right| / \left| Q_{out,r} \frac{dR_{rad}}{dt} \right| >> 1, \label{eq:qr}
\end{equation}

\begin{figure} 
\includegraphics[scale=.65]{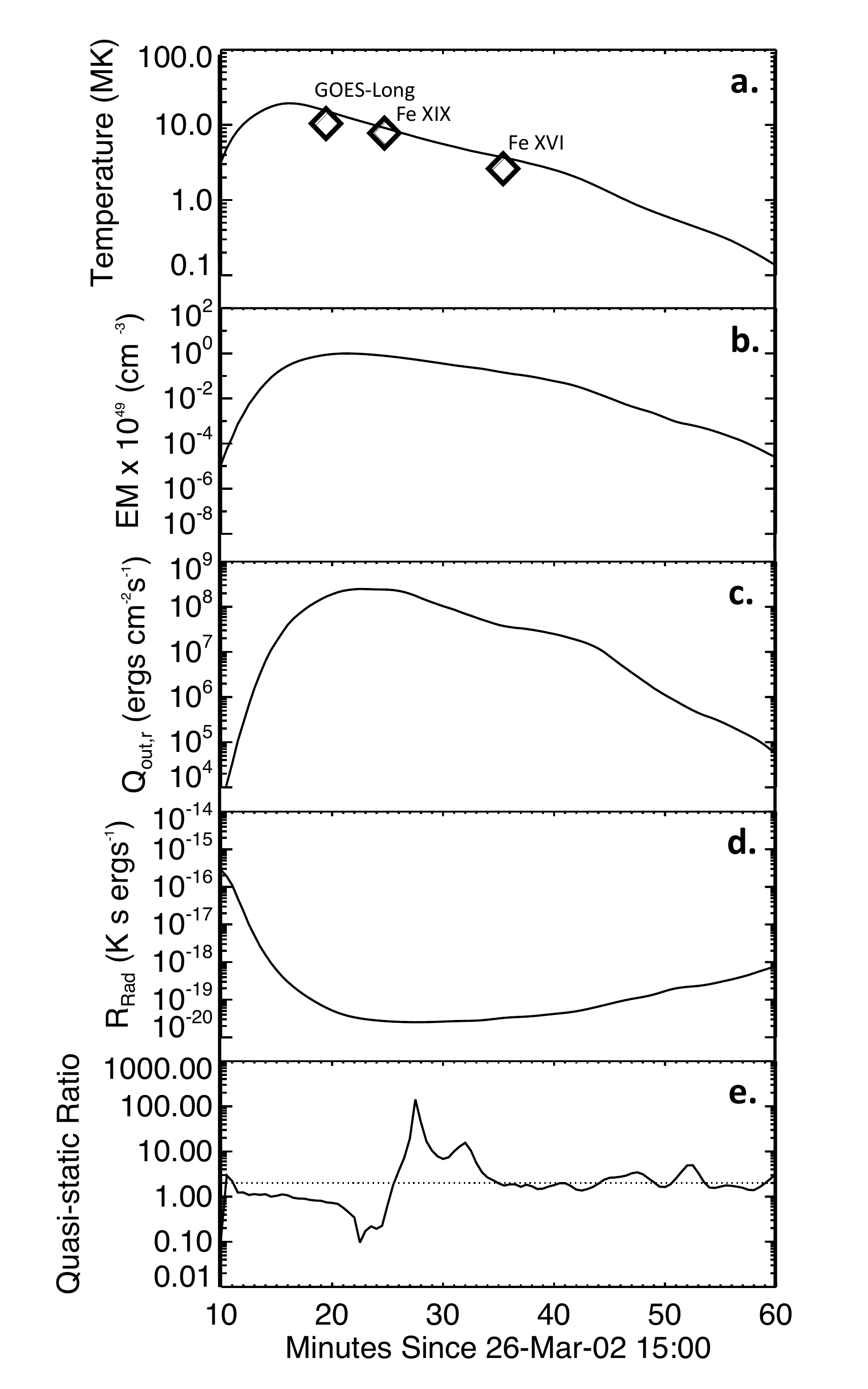}
\caption{Panels a-c show the temperature, emission measure and radiative loss found for a canonical C-class solar flare by \cite{raftery2009multi}.  These values are used to compute $R_{rad}$ and the Quasi-static Ratio shown in Panels d and e.  A Quasi-static Ratio much greater than 1 implies $R_{rad}$ evolves quasi-statically with respect to $Q_{out,r}$.  The diamonds in Panel a correspond with when the indicated measurements peaked.   }
\label{fig:canonical_flare}
\end{figure}

\noindent{}where the term on the left hand side of Equation (\ref{eq:qr})  is the Quasi-static Ratio.  The validity of using the quasi-static approximation is explored by expanding the analysis of \cite{raftery2009multi}, who report the evolution of temperature, emission measure and energy loss for a canonical C-class solar flare predicted by EBTEL and constrained by multi-wavelength observations.  These quantities are shown in Panels a-c of Figure \ref{fig:canonical_flare}.  For this flare, EBTEL predicts radiative cooling dominates over conductive cooling after 15:24.  Panel d reports $R_{rad}$ calculated from the values in Panels a and b using Equation (\ref{r_eff}), and  Panel e shows the Quasi-static Ratio  computed from the values in the preceding panels with a solid line; the Ratio value of 2 is over-plotted with a dotted line.  Near the time when conductive cooling begins to dominate, the Quasi-static Ratio rapidly increases from values near 0.1 to values exceeding 10 near 15:25 and remains at or above 10 until approximately 15:33, at which point it decreases to values near 2 for the duration of the flare.  Panel a also shows with diamonds the times where the GOES-Long band and Fe XIX and Fe XVI emissions reach their peak intensity.  From Panels a and d, it is concluded that it is reasonable to treat $R_{rad}$ as being quasi-static with respect to $Q_{out,r}$ during periods when hot EUV emission lines peak.     

Now, when $R_{rad}$ is quasi-static with respect to $Q_{rad}$ , it can be factored out of the derivative and the substitution $RC=\tau_{cool}$ can be made, where $\tau_{cool}$ is the radiative cooling time of the loop,
\begin{equation}
Q_{in}-\tau_{cool}\frac{dQ_{out,r}}{dt}=Q_{out,r}. \label{lpf_thermal}
\end{equation}

If the plasma temperature is near the formation temperature of a bright EUV emission line with intensity, $I_C$, then $Q_{out,r} = \alpha I_C$, where $\alpha$ is a proportionality constant.  Applying this to Equation (\ref{lpf_thermal}) yields,

\begin{equation}
\frac{Q_{in}}{\alpha}-\tau_{cool,C}\frac{d I_C}{dt}= I_C \mbox{     when   } T_C-\delta T<T< T_C+\delta T.  \label{lpf_thermal_final}
\end{equation}

\noindent{}where $T_C$ is the peak formation temperature of $I_C$, and the constraint, $ T_C-\delta T<T< T_C+\delta T$, requires $\tau$ to be modified to $\tau_{cool,C}$, the cooling time constant of the ions emitting $I_C$.

We use Equation (\ref{lpf_thermal_final}) to motivate the form of an empirical relationship, referred to as LETM, between the intensity of some hotter emission line, $I_H$, and some cooler emission line $I_C$.  Specifically, the hotter emission line flare irradiance, $I_{H}$, is first found by subtracting off its pre-flare background.  This value is then normalized by its maximum value and input to LETM:

\begin{equation}
\frac{I_{H}}{max(I_H)}-\tau\frac{ I'_{C} }{dt}=I'_{C}. \label{eqn_lpfe_i}
\end{equation}

\noindent{}Here, $I'_C$ is the modeled output intensity of the cooler emission line and is linearly proportional to the actual flare irradiance, $I_C$.  $\tau$ is taken to be equal to the time difference between the hotter and cooler emission peak values because this value was found to result in $I'_C$ and $I_C$ peaking simultaneously and having comparable widths.  We do not fully understand why Equation (\ref{eqn_lpfe_i}) describes the observations, specifically the inferred proportionality between $I_H$ and $Q_{in}$, but we will discuss the implications of the LETM relation to flare physics in Section \ref{sec:disc} after the observational evidence is presented.

It is shown in Section \ref{sec:obs} that recursive applications of Equation (\ref{eqn_lpfe_i}) can improve model-measurement linear correlations for cases where the formation temperature differences between $I_H$ and $I_C$ are relatively large.  In these cases, Equation (\ref{eqn_lpfe_i}), is first solved for $I'_C$.  $I'_C$ is then used as an $input$ to Equation (\ref{eqn_lpfe_i}) (in place of $I_H$) whose output is compared with the cooler emission irradiance measurements.  In this paper, Equation (\ref{eqn_lpfe_i}) is not solved recursively more than twice.  Two independently determined values for $\tau$ are used, one for each application of Equation (\ref{eqn_lpfe_i}), and found using the aforementioned method of differencing the measured peak intensity times. 

The shorthand notation ``$L(I_H)$'' is used in place of Equation (\ref{eqn_lpfe_i}) where $H$ is replaced with the particular species being filtered.  For twice recursive applications of Equation (\ref{eqn_lpfe_i}), ``$L(L(I_H))$'' is used.

\begin{figure} 
\includegraphics[scale=.5]{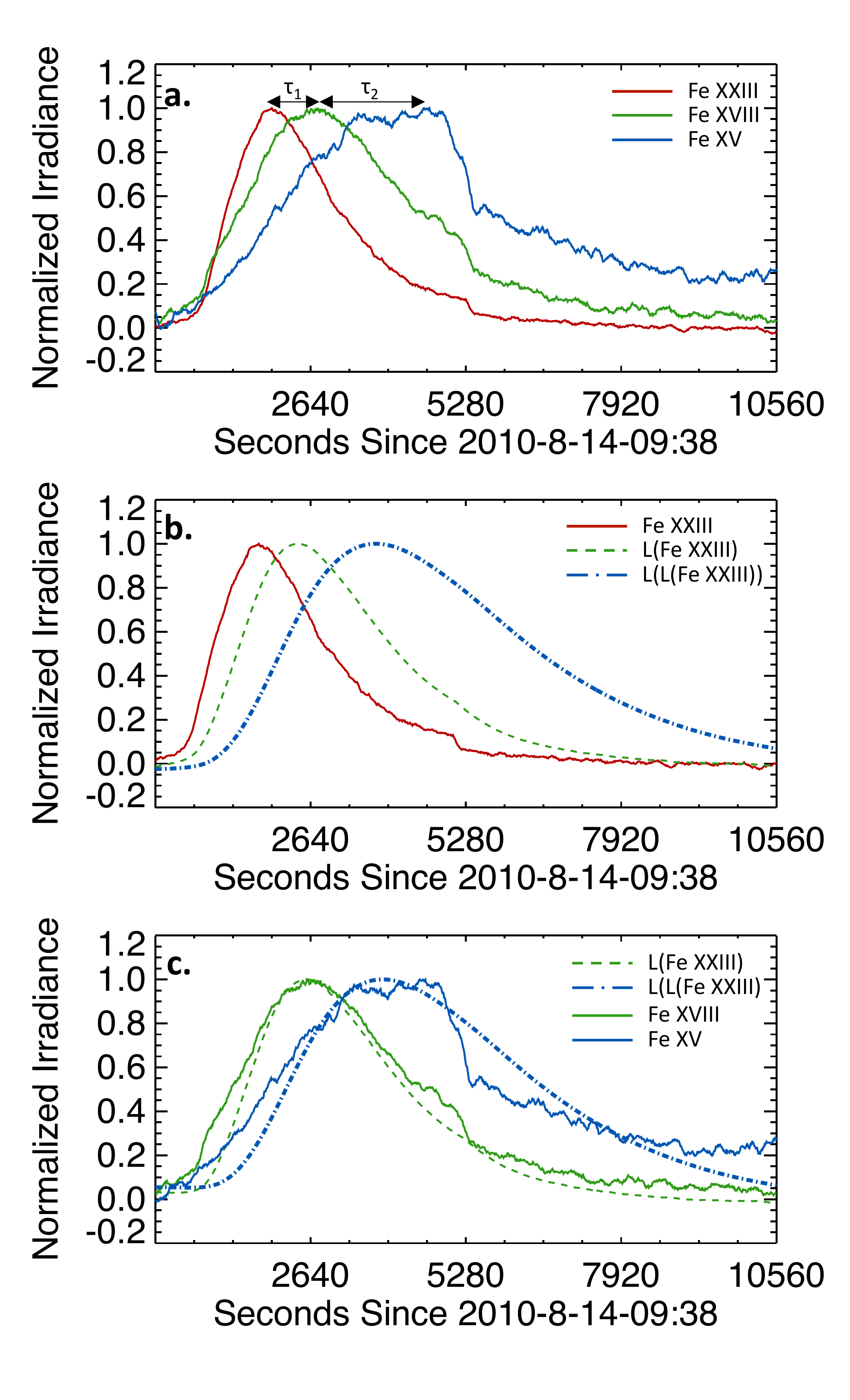}
\caption{An example application of the LETM method.  a) Measured light curves for Fe XXIII, Fe XVIII and Fe XV.  b)  The Fe XXIII light curve is shown with the solution of single ( L(Fe XXIII) ) and twice recursive applications ( L(L(Fe XXIII))  )  of Equation \ref{eqn_lpfe_i}.  c)  The Fe XVIII light curve and L(Fe XXIII) solution are highly correlated, as are the Fe XV light curve L(L(Fe XXIII)) solution.   }
\label{fig:stepwise}
\end{figure}

This method is illustrated through the following example, which applies LETM to a C4.4 flare occurring on 14 August 2010.  Figure \ref{fig:stepwise}a shows normalized EVE light curves for the Fe XXIII (13.31 nm), Fe XVIII (9.4 nm) and Fe XV (28.4 nm) lines measured by SDO EVE.  These data are used to determine $\tau_1$ and $\tau_2$ used in the next steps, and Figure \ref{fig:stepwise}a shows schematically how the time-constants are defined.  The Fe XXIII line irradiance is input into Equation \ref{eqn_lpfe_i} as $I_H$, and it is solved for $I_C'$ with time-constant $\tau_1$.  This first solution is referred to as L(FeXXIII) and shown in Figure \ref{fig:stepwise}b with a green dashed curve.  L(FeXXIII) is this input into Equation  \ref{eqn_lpfe_i} as $I_H$,  and it is solved with time-constant $\tau_2$.  This solution is referred to as L(L(FeXXIII)), and is shown with a dashed-dotted line in Figure \ref{fig:stepwise}b.  Figure \ref{fig:stepwise}b also shows the Fe XXIII light curve for comparison.  Fe XVIII and L(FeXXIII) are highly correlated, as are Fe XV and L(L(FeXXIII)).  This is demonstrated in Figure \ref{fig:stepwise}c, where the four light curves are over plotted.

\section{Data Sources} \label{sec:data_methods}

We use EUV flare observations made by the SDO EVE Multiple EUV Grating Spectrograph (MEGS)-A channel which made near continuous full disk solar irradiance measurements in the 6-35 nm range at 0.1 nm resolution and 10 s time cadence from 1 May 2010 through 26 May 2014.  For this study, we use the Level 2 Version 5 \emph{Spectrum} data product, from which we derive the line irradiances for the Ni XVIII emission line at 29.22 nm as well as 10 emission lines from different ionization states of Fe with peak formation temperatures ranging from 0.89 MK to 14.12 MK using line center and width values reported in the EVE Flare Atlas in \cite{hock2012thesis}.  The ionization states and line center wavelengths are reported for all 11 lines in Section \ref{sec:perf_sum}. 
The selected lines span a broad range of flare temperatures, have a measurable flare contribution, and are spectrally pure with the specified lines contributing at least 85\% of the flare irradiance across the line wavelength range.  

Flares are identified using the GOES XRS Event List that is retrieved programmatically using the SolarSoft (\cite{freeland1998data}) routine, pr\_gev.pro.

\section{Results} \label{sec:obs}
\subsection{Model and Measurement Comparison for 12 Flares} \label{sec:fig12}

\begin{figure} 
\includegraphics[scale=.38]{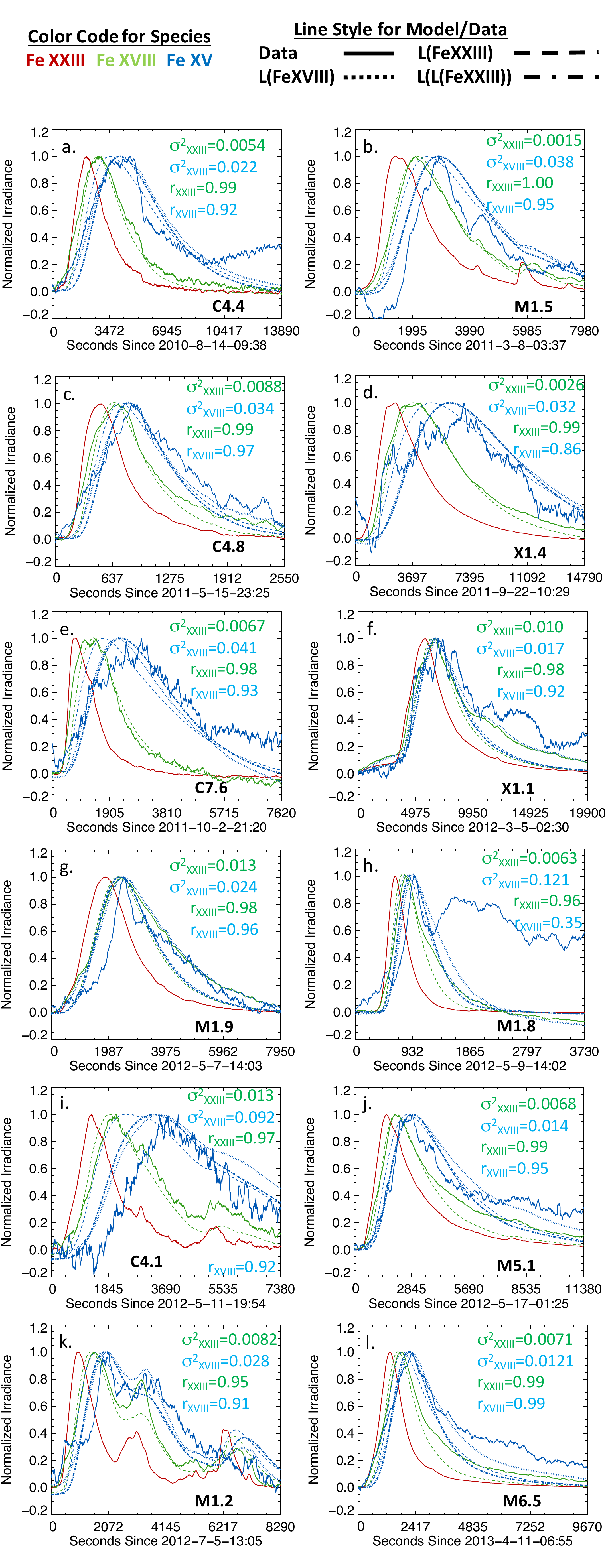}
\caption{Comparisons of Equation (\ref{eqn_lpfe_i}) predictions with observations for 12 lines.  The color represents the emission line and the line-style represents the model or data shown.  Solid lines correspond with EVE measurements for the emission line whereas the other four line-styles represent models for the corresponding line.   }
\label{fig:ex_plate}
\end{figure}

Examples of Equation (\ref{eqn_lpfe_i}) applied to 12 flares are shown in Figure \ref{fig:ex_plate}.   In each panel, measurements of the 13.3 nm Fe XXIII, 9.4 nm Fe XVIII and 28.4 nm Fe XV light curves are shown in solid red, green and blue, respectively.  L(Fe XXIII) predictions of Fe XVIII and Fe XV are shown in dashed green and blue, respectively.  L(L(Fe XXIII)) predictions of Fe XV are shown in dot-dashed blue.  The time difference between the measured Fe XXIII and Fe XVIII peaks are used for the first LETM time-constant ($\tau$) and the time difference between the measured Fe XVIII and Fe XV are used for the second LETM time-constant ($\tau$).  Finally, L(Fe XVIII) predictions of Fe XV are shown with dotted blue curves.  All model outputs and measurements are normalized by their respective maximum values to focus on modeling the irradiance time evolution rather than the absolute magnitude.  

Model-measurement variances are shown for the Fe XXIII- and Fe XVIII-based predictions of Fe XVIII and Fe XV designated by $\sigma ^2_{XXIII}$ and $\sigma ^2_{XVIII}$, respectively.  For these same measurement-model pairs, Pearson correlation coefficients are also given in each panel and designated by $r_{XXIII}$ and $r_{XVIII}$, respectively.  The variances have units of normalized irradiance and, therefore, are unitless. The Fe XV measurements are noisier than the other two emission lines, contributing to the larger variances.  Both the correlation coefficients and variances are useful for relative comparison between the model-measurement pairs, although the variances are more sensitive to model-measurement disagreement than the correlation coefficients.  See panel 3l for example, where the correlation coefficients for both model-measurement pairs equals 0.99 but the $\sigma_{XXIII}$ variance is half as large as the $\sigma_{XVIII}$ variance, indicating that, for this particular flare, Fe XXIII is a better predictor of Fe XVIII than Fe XVIII is of Fe XV. 

For the flares shown in Figure \ref{fig:ex_plate}, there is a very high degree of linear correlation between L(Fe XXIII) and Fe XVIII, with Pearson correlation coefficients greater than 0.99 in half of the examples shown and a minimum value of 0.95 for all twelve examples.  
Further, both L(Fe XVIII) and L(L(Fe XXII)) capture the delay and broadening of the Fe XV light curves for the examples shown, but the agreement tends to be worse than that seen between Fe XVIII and L(Fe XXIII).  Some dissimilarities between L(Fe XVIII) and L(L(Fe XXIII)) and Fe XV are due to flare phases that appear in Fe XV but not the two hotter lines.  For example, the peak seen in Fe XV near 2000 s in Panel d has the signature of an impulsive flare component, and the large late gradual rise after the initial gradual phase in Panels a and h are indicative of the EUV Late Phase (\cite{woods2011new}).  Finally, L(Fe XXIII) is a worse predictor of Fe XV than L(L(Fe XXIII)), specifically, L(Fe XXIII) tends to predict the rise phase and peak too early while more accurately predicting the decaying phase.  

From Figure \ref{fig:ex_plate}, we see Equation (\ref{eqn_lpfe_i}) is capable of propagating the structures that appear in the hot emission lines to the cooler emission lines.  An obvious example of this is seen in Panel \ref{fig:ex_plate}k, where the distinct double peak evident in the Fe XXIII is broadened and merged by the models which also reduce the relative difference between the peaks in a manner consistent with the data.  Even with this complex evolution, the correlation coefficient for the Fe XXIII (Fe XVIII) model of  Fe XVIII (Fe XV) is 0.95 (0.91).

Note that Equation (\ref{eqn_lpfe_i}) also relates emission lines with substantially different peak formation temperatures.  Fe XVIII forms at temperatures that are 37\% cooler than Fe XXIII, and Fe XV forms at temperatures that are 72\% and 82\% cooler than Fe XVIII and Fe XXIII, respectively.  The reasonable agreement between the Fe XV and Fe XVIII predicted light curves derived from Fe XVIII and Fe XXIII measurements, respectively, suggests lines can differ by as much as 72\% and reasonable model and measurement agreement can be found using Equation (\ref{eqn_lpfe_i}).

\subsection{Characterized Model Performance for Additional Lines and Flares} \label{sec:perf_sum}

The analysis is expanded to include the 11 emission lines listed in Table \ref{tab:lpf_perf}.  The species, center wavelength and formation temperature logarithm are given in Columns A-C, respectively.  All flares observed by the SDO EVE, MEGS-A channel are considered that also have a GOES XRS magnitude C.4 or greater, and MEGS-A line measurements with a SNR $>$  3 for the line to be modeled.  The number of flares considered for a given emission line are listed in Column H of Table \ref{tab:lpf_perf}.  For each line, we predicted flare light curves using the same three model variants used in Section \ref{sec:fig12}.  The model quality for each emission line is characterized using the $N_{flr}/e$ smallest model-measurement variance, $\sigma_{e}^2$.  This choice is arbitrary, but provides a value for comparison that is representative of the entire population.  These values are unitless because the model outputs and measurements are normalized by their maximum values.  
  $\sigma_{e}^2$ values for the three models considered are shown in Columns E-G.  In Column D, we give the $\sigma_{e}^2$ value found between measurements and \emph{unmodified} normalized Fe XXIII light curves; these values are representative of models like FISM that do not broaden nor delay the input proxy.   The values in Column J are discussed in Section \ref{sec:disc}.   

\begin{table}
\small
\linespread{.5}\selectfont
\centering
  \caption{Characteristic model-measurement variances for 10 lines.  The number of flares considered for each line is given in Column H.  The F-Test critical values for the $F(N_{flr},N_{flr})$ distribution at the 1\% significance level are given in Column I.  The characteristic variance is the $N_{flr}/e$ smallest variance of the flares analyzed and reported for the 3 model variants, L(XXIII), L(XVIII) and L((XXIII)) discussed in the text.  For reference, variances are given in Column D which correspond with differences found using normalized Fe XXIII light curves.  Comparing Columns E-G with Column D shows better model-measurement agreement for all three model variants.  However, there is a marked increase in model-measurement disagreement for the the model variants for lines cooler than Fe XV.  Column J gives the correlations between measured $\tau$ values and those predicted by a method discussed in Section \ref{sec:disc}.   }
    \label{tab:lpf_perf}
   \begin{tabular}{cccccccccc}
     \hline 
             A&B&C&D&E&F&G&H&I&J\\
             Species & $\lambda$ (nm) & $Log(T_{max})$  & $ \sigma^2_{e,Fe23}$& $\sigma^2_{e,L(Fe23)}$& $\sigma^2_{e,L(Fe18)}$ & $\sigma^2_{e,L(L(Fe23))}$&$N_{flr}$ & $F_{crit}$&$r_\tau$\\
     \hline 
      FeXXIII&13.31&7.15&NA&NA&NA&NA&1160&NA&NA\\
      FeXXI&12.9&7.1&0.0056&0.0034&NA&NA&1156&1.15&0.53\\
      FeXIX&10.85&7.0&0.0398&0.0115&NA&NA&1155&1.15&0.77\\
      FeXVIII&9.4&6.95&0.0762&0.0202&NA&NA&1143&1.15&NA\\
      NiXVIII&29.22&6.9&0.1436&0.0836&0.0592&0.0841&234&1.36&0.93\\
      FeXVI&33.55&6.8&0.1811&0.0918&0.0625&0.0799&313&1.30&0.93\\
      FeXV&28.43&6.4&0.2139&0.1168&0.0884&0.0992&464&1.24&0.79\\
      FeXIII&20.12&6.3&0.1534&0.1341&0.1443&0.1366&109&1.57&0.74\\
      FeXII&19.52&6.25&0.157&0.1417&0.1469&0.1483&216&1.37&0.71\\
      FeXI&18.05&6.15&0.185&0.1649&0.1685&0.1712&164&1.44&0.60\\
      FeIX&17.12&5.95&0.1676&0.1492&0.1599&0.1508&135&1.50&0.71\\
     \hline 
    \end{tabular}%
\end{table}

Column I gives the critical value, $F_{crit}$, for the F distribution, $F(N_{flr},N_{flr})$, at the 1\% level for the number of flares in each row.  These can be used to evaluate an F Test for statistically significant differences in the variances reported in each row (\cite{von2001statistical}).  If the ratio of two variances, with the smaller variance in the denominator, is found to be greater than or equal to $F_{crit}$, the differences in the variances are statistically significant in 99\% of cases, and it can be concluded that the model with the smaller variance is an improvement over the model with the larger variance.  

Comparing Columns D and E, we find that L(FeXXIII) has markedly smaller (better) model-measurement variances than the unmodified Fe XXIII light curves for Fe XV and hotter formed lines.  For lines with Log formation temperatures between 6.4 and 6.9, it is evident that L(FeXVIII) and L(L(FeXXIII)) provide further improvement over L(FeXXIII).  For lines that are cooler than Fe XV, the variances in Columns E-G are not significantly smaller, as determined by the F-test, than the reference variances in Column D.  The breakdown of the empirical relation at these cooler lines may be because the formation temperature difference between the lines has become too large, and improvement may be found by (further) recursive application of Equation (\ref{eqn_lpfe_i}) to Fe XV (or Fe XXIII).  On the other hand, improvement for the cooler lines may not be possible if the discrepancy is due to heating processes occurring later in the flare that heat the plasma to the formation temperatures of the cooler lines but not to the temperature of input line.  Or, the decreased model performance at these cooler lines may be due to the coronal dimming and impulsive flare phases which are more common in the cooler emissions (\cite{woods2011new}).  Further, Figure \ref{fig:canonical_flare}e shows the quasi-static ratio decreases as the flare cools, implying the LETM approximation for flare cooling will begin to break down later in the gradual phase, when these cooler lines reach their peak formation temperatures.  Finally, we note that the Columns E-G show statistically significant smaller variances with respect to Column D for predictions of Ni XVIII from the two Fe lines; this suggests the empirical relation given by Equation (\ref{eqn_lpfe_i}) holds when comparing different species.  Although Table \ref{tab:lpf_perf} is useful for understanding how Equation (\ref{eqn_lpfe_i}) can improve flare irradiance modeling, detailed analysis of light curves from each of these lines is necessary to draw conclusions about the underlying physical processes.


\subsection{Measured $\tau$ Time Constants for Select Lines} \label{sec:tau}

\begin{figure} 
\includegraphics[scale=.5]{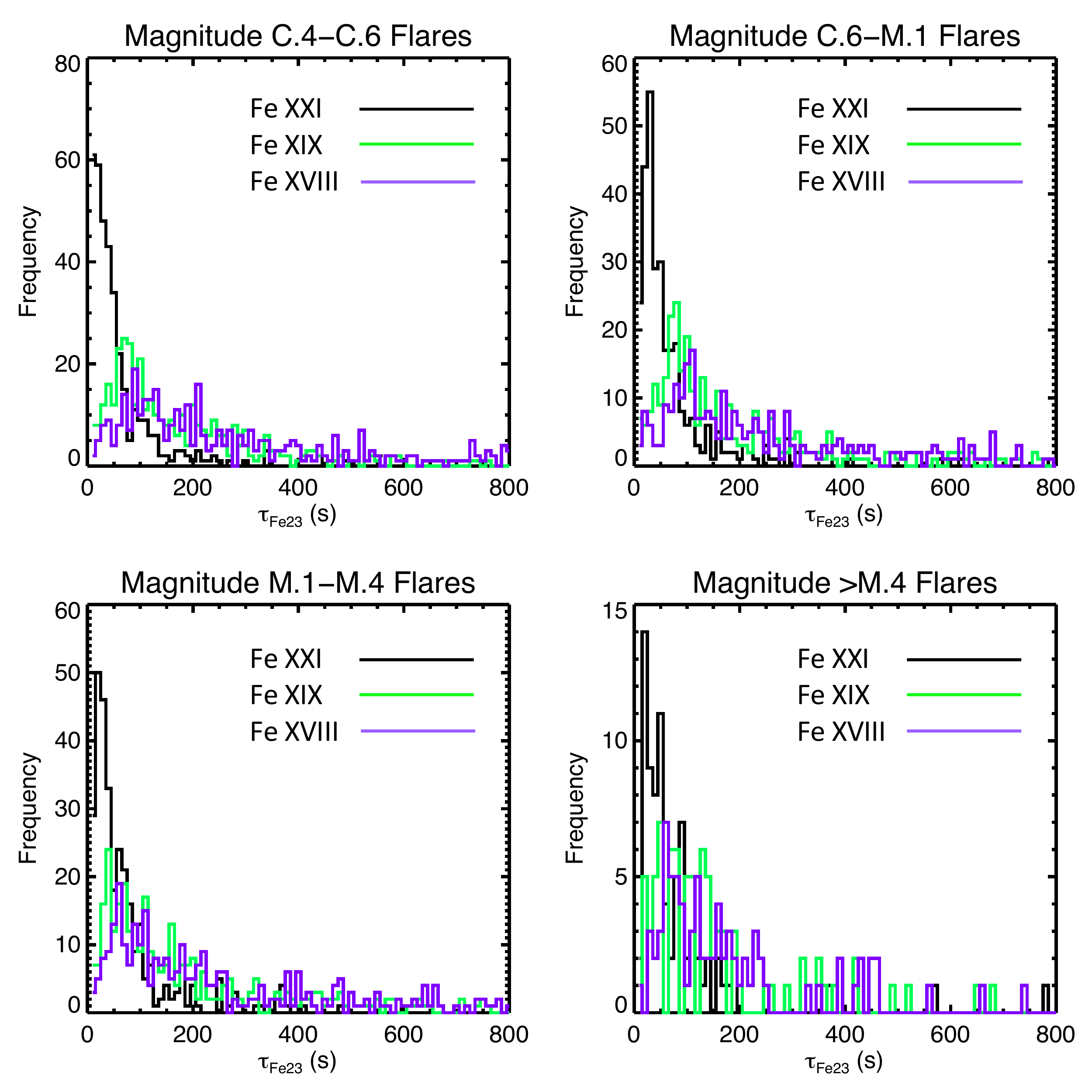}
\caption{Histograms of the \deleted{LPF}\added{LETM} time constant, $\tau_{Fe23}$, for the L(FeXXIII) model for flares grouped according to GOES XRS magnitude range shown at the top of each panel.  $\tau_{Fe23}$ is the \deleted{filter}\added{LETM} time constant, and taken to be the measured time difference between the time of peak intensity of the indicated line and the Fe XXIII line.   }
\label{fig:tau_fe23}
\end{figure}

Given that Equation \ref{eqn_lpfe_i} is constrained by only the time constant, $\tau$, the observed values of $\tau$ warrant further consideration.  Histograms of the time constants, $\tau_{Fe23}$, for the L(FeXXIII) model are shown in Figure \ref{fig:tau_fe23} for the three hottest lines with formation temperatures below that of Fe XXIII.  To control for any dependence on flare magnitude, the total set of flares are partitioned into four groups based on the flare soft x-ray magnitude.  As discussed in Section \ref{sec:methods},  $\tau_{Fe23}$ values are found by differencing the time of peak intensity for a particular line and the time of peak intensity for the Fe XXIII line.    

Figure \ref{fig:tau_fe23}  shows that the distributions are sharply peaked near for the hottest emission shown, Fe XXI, and the distribution broadens with decreasing formation temperature for all four flare groups.  However,  for the cooler Fe XIX line, $\tau_{Fe23}$ peaks at 70, 80, and 40 seconds for the magnitude C.4-C.6, C.6-M.1 and M.1-M.4 groups, respectively, indicating a marked increase in the flare cooling rate for the M.1-M.4 groups of flares.  Note, it is difficult to identify a definitive distribution peak for the Fe XIX line in the $>$M.4 group.  The right-handed half-widths at half-maximum are approximately 30, 40, and 100 seconds for the Fe XXI, Fe XIX and Fe XVIII emission lines, respectively.  For ions cooler than Fe XVIII (not shown), the distribution peak becomes increasingly less distinguishable from the background.  This may simply be because the number of total flares with adequate SNR is reduced by an approximate factor of 4 (see Column H of Table \ref{tab:lpf_perf}), or it may be indicative of cooling processes becoming less consistent for lower temperatures.  

\begin{figure} 
\includegraphics[scale=.4]{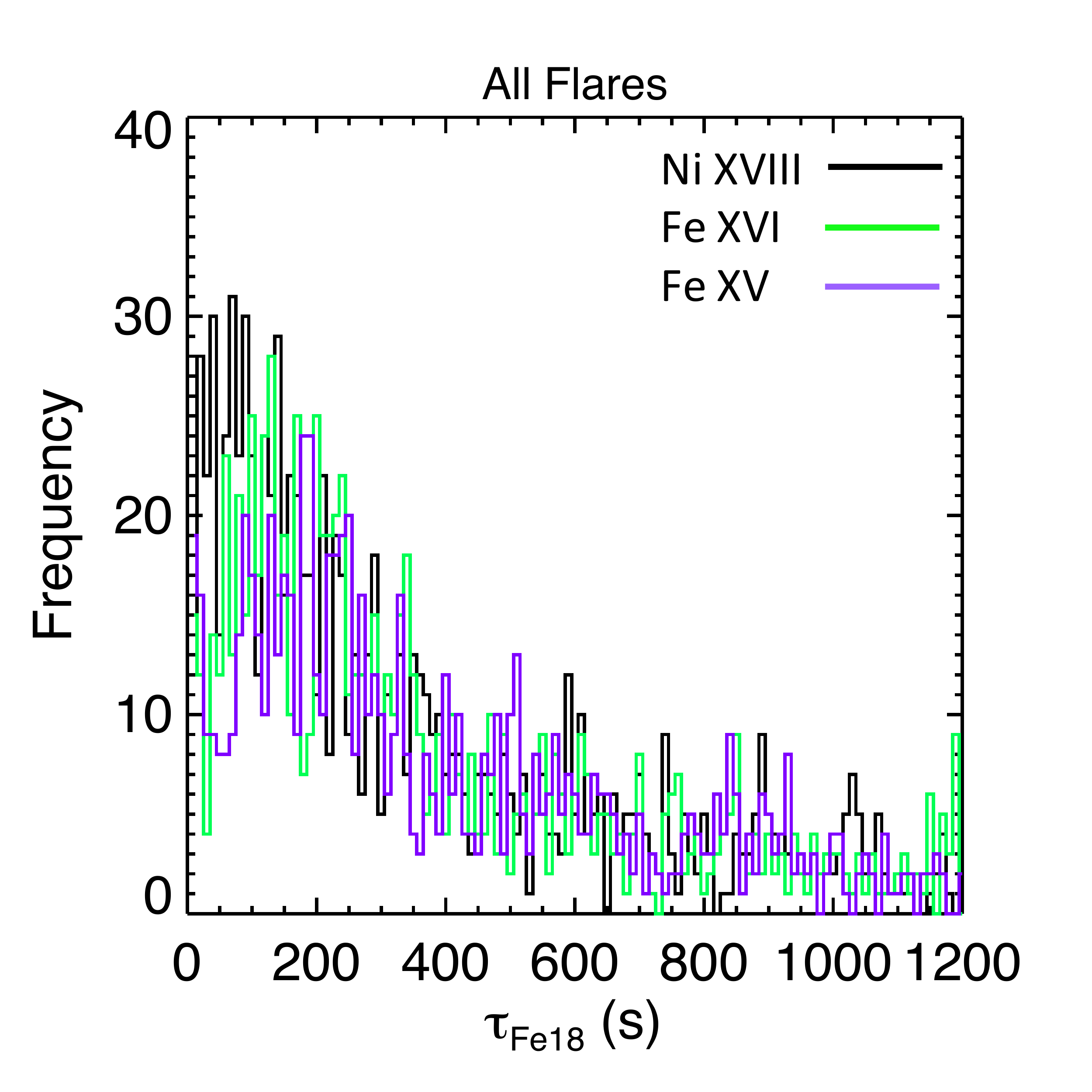}
\caption{Similar to Figure \ref{fig:tau_fe23} but for the L(FeXVIII) time-constants which are measured with respect to the Fe XVIII time of peak intensity.  Due to the smaller sample size, all flares are considered in a single histogram rather than being sorted according to flare magnitude.   }
\label{fig:tau_fe18}
\end{figure}

Figure \ref{fig:tau_fe18} shows the L(FeXVIII) time constants, $\tau_{Fe18}$, for the three hottest lines with formation temperatures below that of Fe XVIII.  In this case, the time constants are measured with respect to the time of peak intensity of Fe XVIII emission line.  Due to the smaller sample size, flares are not sorted according to magnitude.  As would be expected for a cooling plasma with a cooling rate that decreases with time, the values of $\tau_{Fe18}$ increase with decreasing line formation temperature.  Here, we see the peak distributions of $\tau_{Fe18}$ are equal to approximately 75, 125 and 200 seconds with right-handed half-widths at half-maximum of approximately 200 seconds for the Ni XVIII, Fe XVI and Fe XV emission lines, respectively.

%
%

\section{Discussion} \label{sec:disc}

\subsection{Implications for Flare Irradiance Modeling}

In terms of flare irradiance modeling, Equation (\ref{eqn_lpfe_i}) can improve estimates of spectral energy deposited in planetary upper atmospheres by solar flares.  The flare in Figure \ref{fig:ex_plate}l is used as an example to quantify the impact of EUV flare irradiance model errors in predicting the relative energy deposited.  To focus on errors from the irradiance time evolution, the peak line irradiance is assumed to be known or determined independently.  The FISM model driven with Fe XXIII as a flare proxy (instead of the nominal XRS-Long flare proxy) is used for comparison.  FISM predicts the Fe XVIII integrated energy to be 63\%, whereas L(FeXXIII) predicts 82\% of measured for this flare; and FISM predicts the Fe XVIII line to peak 510 seconds earlier than measured compared to the 130 second early peak predicted by L(FeXXIII).  For the Fe XV emission line, FISM predicts the integrated energy to be 48\%, whereas L(L(FeXXIII)) predicts it to be 77\% of measured; and FISM predicts the Fe XV line to peak 1030 seconds early compared to the 190 second early prediction of L(L(FeXXIII)).  It is evident that Equation (\ref{eqn_lpfe_i})  can result in a significant improvement of both the flare peak time and total energy deposited using the unmodified Fe XXIII measurements as a proxy for the flare time evolution.    

It is important to note that LETM inherently assumes all heating occurs during the formation of the Fe XXIII when there may be ongoing heating that heats plasma to formation temperatures of cooler lines, but not to temperatures where Fe XXIII peaks.  This will result in a tendency for LETM to underestimate the irradiance of cooler lines, especially in the declining phase of the light curve.  This can be seen in a number of the examples shown in Figure 3, and may explain why LETM under predicts the integrated irradiance in the above example.

With regard to how this empirical model compares with previous studies that use the physics-based EBTEL model to estimate the time evolution of EUV flare light curves, the empirical model of Equation (\ref{eqn_lpfe_i}) can reproduce emission line light curves with significantly more accuracy than has been shown with EBTEL, and does so using fewer inputs and higher computational efficiency.  To fully appreciate this, the reader is encouraged to compare the EBTEL flare light curve modeling examples listed in the second-to-last paragraph of Section \ref{sec:intro} with those shown in Figure \ref{fig:ex_plate}. 

Although Equation (\ref{eqn_lpfe_i}) shows promise for significantly improving flare irradiance models, it is limited by the current inability to independently predict $\tau$ and the typical unavailability of full-disk integrated 13.3 nm Fe XXIII emission line measurements.  Section \ref{sec:tau} shows that the $\tau$ distributions have a definitive peak that broadens with decreasing formation temperature. The broad distribution of $\tau$ values likely precludes the use of the mean $\tau$ value as a static model input parameter.  Since $\tau$ is determined from the times when the two lines reach their peak intensities and the peak intensity should occur at the species peak formation temperature (for a given density), it may be possible to predict $\tau$ from the DEM or possibly more rudimentary isothermal temperature and emission measure estimates if they are known independently.  

\begin{figure} 
\includegraphics[scale=.6]{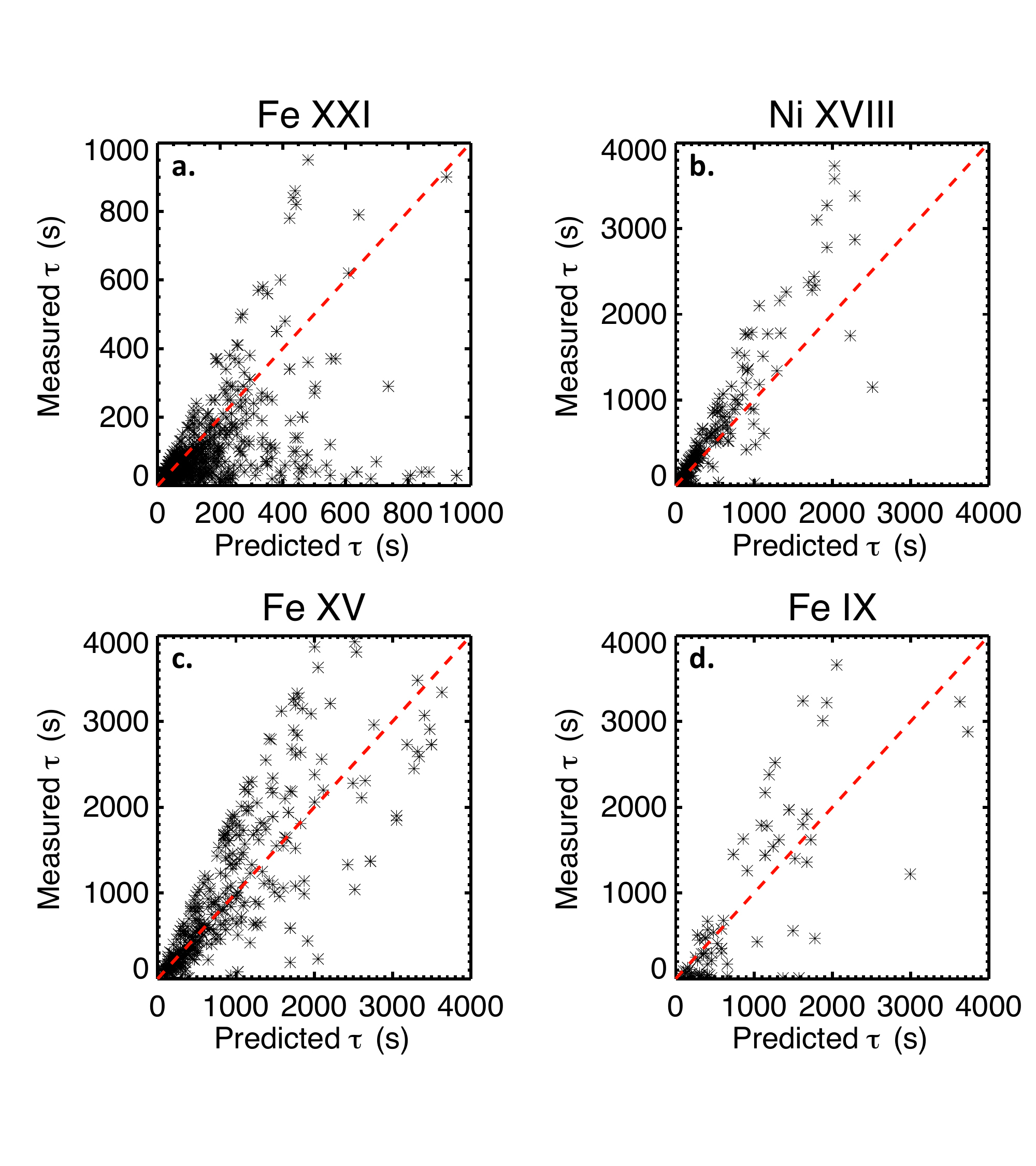}
\caption{Predicted versus measured $\tau$ values for four example emission lines.  $\tau$ predictions were found using Fe XXIII and Fe XVIII peak formation temperatures and peak flare times to estimate a linear cooling rate.  The cooling rate and the peak formation temperatures of the respective lines in the four panels shown are then used to predict $\tau$.  The 1:1 line is shown in dashed red. }
\label{fig:tau_scat}
\end{figure}

To this end, the possibility of empirically estimating $\tau$ by using a select number of emission lines is investigated.  By using 7 Fe lines with formation temperatures between 2 MK and 15.8 MK, \cite{ryan2013decay} found a linear relationship between the time of an emission line's peak intensity and its peak formation temperature (i.e. the cooling rate) during the majority of the 72 flares that they considered.  We attempt to reproduce this result with just two lines by first finding linear estimates of the cooling rate from the Fe XXIII and Fe XVIII emission peaks, and then using the cooling rate estimate to predict when a different emission line will peak (i.e. the $\tau$ value for that line).  We report the correlation coefficients for the lines considered in Column J of Table \ref{tab:lpf_perf}; and scatterplots of the measured and predicted $\tau$ values for 4 example emission lines are shown in Figure \ref{fig:tau_scat}.  Although the correlations in Table \ref{tab:lpf_perf} are strong-to-moderate, there is significant scatter in the 4 example lines evident in Figure \ref{fig:tau_scat}, indicating that this simple method of estimating $\tau$ with two line emissions is inaccurate.  Note that the linearity and correlation improves for emissions with a formation temperature closer to that of Fe XVIII simply because Fe XVIII is used to estimate the cooling rate.  This explains the good predictions of $\tau$ for Ni XVIII in Figure \ref{fig:tau_scat}b compared to the other examples.  One likely source of error in this analysis is that the line intensity is also a function of EM (or DEM), which also evolves significantly during a flare.  Improvements on this method may be possible by incorporating time-dependent EM or DEM estimates.

Flare temperature and EM estimates are available through a number of methods for flare measurements made over the past 30 years, and new operational EUV instruments such as EXIS and the Solar Ultraviolet Imager (SUVI) (\cite{martanez2010novel}), both onboard the GOES-R series satellites, will maintain and improve this capability in the coming decades.  For example, GOES XRS isothermal flare temperature and EM estimates (\cite{thomas1985expressions}) may be suitable for the hottest EUV emission lines, but XRS becomes insensitive to cooler ($<$10 MK)  flare temperatures for moderate (B class) background soft x-ray activity (\cite{feldman1996electron}), limiting its applicability for estimating when lines with cooler formation temperatures reach peak intensity.  Cooler temperature sensitivity may be attained with broadband measurements that are sensitive to cooler plasma such as those on the EUV Spectral Photometer (ESP) channel on SDO EVE (\cite{didkovsky2009euv}) and the Mars Atmosphere and Volatile EvolutioN (MAVEN) EUV Monitor (EUVM) (\cite{eparvier2015extreme}) or EUV line measurements made by the MEGS-B channel on SDO EVE.  Alternatively, peak emission times measured by imagers such as SDO Atmospheric Imaging Assembly (AIA) (\cite{lemen2012atmospheric}) may be combined with full disk irradiance measurements, such as those from the SDO EVE MEGS B channel, to estimate $\tau$.  This approach may be especially viable with the imminent availability of measurements from the GOES SUVI imager and EXIS full disk irradiance sensors that will observe a combined 11 different emission lines in addition to the two traditional XRS bands.  Regarding alternatives to the Fe XXIII 13.3 nm EVE observations, XRS-Long may be a suitable alternative because observations suggest that it evolves similarly to Fe XXIII (e.g. \cite{woods2011new}).  

Because robust methods of independently predicting $\tau$ have yet to be developed, detailed characterization of how Equation  (\ref{eqn_lpfe_i}) improves FISM or other flare irradiance models is premature.  This paper will be followed by a study that investigates methods of independently predicting $\tau$ and applying the empirical findings reported here to research and operational flare irradiance models.

In summary, a model using Equation (\ref{eqn_lpfe_i}) can be implemented in the following straightforward fashion if the timing of the emission line peaks are predicted from independent temperature and emission measure measurements.  First, each model bin is assigned a characteristic peak formation temperature, determined from the flare lines within a given bin. These values are the same for all flares.  Next, line peak time predictions are used to compute a unique $\tau$ value for each model bin.  Bins that contain emissions that are significantly cooler than the input emission are assigned two or more $\tau$ values to allow for iterative application of Equation (\ref{eqn_lpfe_i}).  Finally, using a single hot emission line as an input, Equation (\ref{eqn_lpfe_i}) is computed for each model bin (iteratively for the bins with more than one $\tau$ value) yielding the time evolution of the flare irradiance for that particular bin.

\subsection{Implications for Flare Cooling Diagnostics}

A direct comparison of Equations (\ref{eqn_lpfe_i}) and (\ref{lpf_thermal_final}), supported by the empirical evidence in Figure   \ref{fig:ex_plate}, indicates that $I_H$ is proportional to the energy flux exciting $I_C$ near the formation temperature of $I_C$.
  In other words, \emph{the energy leaving a population of hotter-formed ions is directly proportional to the energy entering a population of cooler-formed ions if the formation temperatures of the two populations are not too different.}  For example, Fe XXIII and Fe XVIII are close enough temperature for this effect to hold, but Fe XXIII and Fe XV are not.  As such, the empirical effect described by Equations (\ref{eqn_lpfe_i}) provides new insight into flare cooling processes.

The approximations of pure radiative cooling and $ T_C-\delta T<T< T_C+\delta T$ may explain features seen in the observations.  In each of the Fe XVIII (green) model-measurement examples of Figure  \ref{fig:ex_plate}, Equation (\ref{eqn_lpfe_i}) under-predicts the irradiance at the initial flare rising phase.  This is what would be expected from a model that omits the conductive cooling expected to exist early on in the flare cooling phase.  Or, these early discrepancies could be because the Quasi-static Ratio is small here, invalidating the factoring of $R_{rad}$ from time derivative in Equation (\ref{eq:unfactored}).  With regard to $ T_C-\delta T<T< T_C+\delta T$, this constraint may explain why Equation (\ref{eqn_lpfe_i}) shows better agreement with data for emission lines that are more similar in temperature.  Alternatively, the higher correlation between Fe XVIII and L(FeXXIII) versus Fe XV and L(FeXVIII) may be because the Quasi-static Ratio is much larger during the formation of Fe XVIII than Fe XV, as suggested by the timing of the large values of the Quasi-static Ratio compared to the timing of the hot line formation shown in Figures \ref{fig:canonical_flare} d and a, leading to the degradation of the quasi-static approximation of $R_{rad}$ with respect to $Q_{out,r}$.

In the literature, the time between peaks of different EUV emission lines are often termed \emph{time-lags} (\cite{benz1999heating},\cite{viall2012evidence}, \cite{lionello2015can} ); these time-lags are equivalent to the value chosen for $\tau$ in Equation (\ref{eqn_lpfe_i}).  Time-lags observed in EUV emissions from cooling coronal loops are not well understood (e.g. \cite{klimchuk2015key}); and existing theory and models fail to reproduce observations (e.g. \cite{qiu2012heating}, \cite{klimchuk2010can}, \cite{lionello2015can}).  Equation (\ref{eqn_lpfe_i}) reproduces light curves for cooler emission lines based solely on the time-lag and may provide an opening to reconcile models with observations.  Although Equation (\ref{lpf_thermal_final}) provides some insight, namely relating the $I_C$ excitation energy with the $I_H$ intensity, more theoretical work is needed to understand the physical significance of Equation (\ref{eqn_lpfe_i}) and its relationship to plasma parameters.  Given that this is a 0D effect with no apparent dependence on flare geometry, 0D physics-based models like EBTEL should be useful for better understanding the underlying causes of the relationship governed by Equation (\ref{eqn_lpfe_i}), and its capabilities and limitations for improving empirical irradiance modeling.

The timing of the heating phase during solar flares is still a topic of active debate and it is an open question as to whether the majority of heating typically occurs during the impulsive phase or whether significant on-going heating occurs throughout the gradual phase, so called decay phase heating (\cite{ryan2013decay}, \cite{qiu2016long}, \cite{lionello2015can}).  \cite{ryan2013decay} demonstrate the continuing relevance of the simple Cargill Model by using it to quantify the magnitude of decay phase heating by comparing the measured and predicted cooling times to estimate the decay phase energy.  This type of analysis has the potential to be improved with Equations (\ref{lpf_thermal_final}) and (\ref{eqn_lpfe_i}).  For example, the divergence between model and measurement at 1275 s in Figure \ref{fig:ex_plate}c may be due to heating that occurrs after the initial heating event that creates the Fe XXIII plasma.  On the other hand, Figure \ref{fig:ex_plate} a-e show that the \emph{information} required to create the Fe XV light curve exists in the Fe XXIII light curve, which returns to background near the Fe XV peak, indicating that any decay-phase heating occurring between the rising phase of the Fe XXIII emission and when the Fe XV emission returns to its background level is small.

\section{Conclusions}

1. For typical EUV solar flare light curves, the LETM equation empirically relates emissions from cooler forming lines to emissions from hotter forming lines when the time constant, $\tau$, is taken to be the time difference between the emission line flare peaks.  This relationship is more apparent between the Fe XXIII and Fe XVIII emission lines than the Fe XVIII and Fe XV emission lines and, hence, breaks down as the formation temperature difference between the emission lines increases. EUV flare light curves from ions with peak formation temperatures ranging from 2 MK to 13 MK can be accurately modeled with a single hot emission line, Fe XXIII, as an input; achieving accuracy that exceeds that found by the more complex physics-based EBTEL model.  \\

\noindent{}2.  Equation (\ref{eqn_lpfe_i}) has a well known numerical solution that can be incorporated into both research and operational flare irradiance empirical models.  There are two impediments to implementing this method into a practical flare irradiance model: 1) the ability to independently predict $\tau$ and 2) the availability of a proxy that evolves similarly to the Fe XXIII 13.3 nm emission line.  The former issue may be solved by using independent flare temperature and EM estimates because $\tau$ is directly related to the temperature evolution of the flare plasma; the latter issue may be addressed with the GOES XRS-Long channel which evolves similarly to the Fe XXIII emission line during flares.  \\

\noindent{}3.  The Cargill Model for flare cooling can be factored into the LETM equation once radiative cooling dominates.  Comparison of empirical observations with the factored Cargill Model indicates that the hotter line emission intensity is proportional to the energy flux exciting the cooler emission for emission pairs with peak formation temperatures that differ by as much as 72\%.  It follows that the LETM relationship can be used as a future diagnostic for flare thermal processes.

\appendix

\section{Thermal Resistance Derivation} \label{ap:a}

The heat transport coefficient for radiation, $h_r$, between the flare loop and ambient is defined as

\begin{equation}
h_r=\frac{Q_{rad}}{A_l(T-T_a)}, \label{h_r}
\end{equation}

\noindent{}where $A_l$ is the surface area of the loop (\cite{ghosh_heat_trans}).  \cite{rosner1978dynamics} showed that the power radiated per unit volume of an optically thin coronal plasma, Q/V, is related to the radiative loss function, $\Lambda(T)$, and the respective electron and proton densities, $n_e$ and $n_H$, by 

\begin{equation}
\frac{Q_{rad}}{V}=\Lambda(T)n_e  n_H \approx \Lambda(T)n_e^2, \label{flare_rad}
\end{equation}

\noindent{}where the volume is taken to be that of the loop and assumed to be constant, and we assumed $n_e \approx n_H$.  Further, we assume the simple form of $\Lambda(T)$ reported from  \cite{rosner1978dynamics}. 

 \begin{equation} 
\Lambda(T)=\frac{\Lambda_0}{T^{\frac{2}{3}}}=\frac{10^{-17.73}}{T^{\frac{2}{3}}}\mbox{ }\frac{\mbox{erg-cm$^3$}}{\mbox{sec}}\mbox{, }10^{6.3}<T<10^7 \label{p_t}
\end{equation}

\noindent{}We can now use (\ref{p_t}) and (\ref{flare_rad}) to rewrite (\ref{h_r}) as

\begin{equation}
h_r=\frac{Vn_e^2\Lambda_0}{A_lT^{2/3}(T-T_a)} \approx \frac{Vn_e^2\Lambda_0}{A_lT^{5/3}} \mbox{\space \space \space \space \space} T >>T_a \label{h_r2}
\end{equation}

The thermal resistance for radiation is defined as 

\begin{equation}
R_{rad}=\frac{1}{h_rA} \label{r_h_def},
\end{equation}

\noindent and is a physical quantity in systems that have a small \emph{Biot Number}, $B_i$, which relates the internal thermal conductivity of system to the convection at the boundary (\cite{ghosh_heat_trans}). For a flare loop of radius, $r_l$, and conductivity $\kappa_0$, 

\begin{equation}
B_i=\frac{r_lh_c}{\kappa_0} \approx 0, \label{b_i}
\end{equation}

\noindent{}where $h_c$ is the heat transport coefficient for convection at the boundary of the flare loop and taken to be zero at the loop/vacuum boundary.  Qualitatively, a small $B_i$ corresponds with a system with negligible temperature gradients perpendicular to the boundary (but within the volume defined by the boundary).  Thus, if we assume this to be the case, 

\begin{equation} 
R_{rad}=\frac{T(t)^{5/3}}{\Lambda_0 n_e^2V},\label{r_eff}
\end{equation}

\noindent{}which is related to the radiated heat according to

\begin{equation}
Q_{rad}=\frac{T}{R_{rad}} \label{q_of_r}
\end{equation}
by combining equations (\ref{h_r}) and (\ref{r_h_def}).


\begin{acknowledgements}

This work has been funded by the NASA MAVEN mission and the NASA/NOAA GOES-R EXIS contract.  The SDO EVE data used in this study is publicly available at http://lasp.colorado.edu/home/eve/data/.  The authors would like to thank Drs. A.R. Jones, R.A. Hock and J.P. Mason for their helpful feedback for earlier versions of this manuscript.  E.M.B.T. would like to thank Dr. N.M. Viall for her feedback on the topic of this paper.

\end{acknowledgements}


\bibliography{swsc}


\end{document}